\documentstyle[preprint,prl,aps]{revtex}
\tighten
\begin{document}
\draft

\preprint{\hbox{JLAB-THY-00-09}}
\title{Realistic Calculation of the $hep$ Astrophysical Factor}
\author{L.E.\ Marcucci$^{({\rm a})}$, R.\ Schiavilla$^{({\rm a,b})}$,
M.\ Viviani$^{({\rm c})}$, A.\ Kievsky$^{({\rm c})}$,
S.\ Rosati$^{({\rm c,d})}$}
\address{$^{({\rm a})}$Department of Physics, Old Dominion University, Norfolk, VA 23529}
\address{$^{({\rm b})}$Jefferson Lab, Newport News, VA 23606}
\address{$^{({\rm c})}$INFN, Sezione di Pisa, I-56100 Pisa, Italy}
\address{$^{({\rm d})}$Department of Physics, University of Pisa, I-56100 Pisa, Italy}
\date{\today}
\maketitle
\begin{abstract}
The astrophysical factor for the proton weak capture on $^3$He
is calculated with correlated-hyperspherical-harmonics bound and
continuum wave functions corresponding to a realistic Hamiltonian
consisting of the Argonne $v_{18}$ two-nucleon and Urbana-IX three-nucleon
interactions.  The nuclear weak charge and current operators
have vector and axial-vector components, that include one- and many-body terms.
All possible multipole transitions connecting any of the $p\,^3$He S- and
P-wave channels to the $^4$He bound state are
considered.  The $S$-factor at a $p\,^3$He center-of-mass
energy of $10$ keV, close to the Gamow-peak energy, is predicted to be
$10.1 \times 10^{-20}$ keV~b, a factor of five
larger than the standard-solar-model value.  The P-wave transitions
are found to be important, contributing about 40 \% of the
calculated $S$-factor.
\end{abstract}

Recently, there has been a revival of interest in the reaction
$^3$He($p$,$e^+ \nu_e$)$^4$He~\cite{BK98}.
This interest has been spurred by the Super-Kamiokande collaboration
measurements of the energy spectrum of electrons recoiling
from scattering with solar neutrinos~\cite{Fuk99}.  At energies larger
than 14 MeV more recoil electrons have been observed than
expected on the basis of standard-solar-model (SSM) predictions~\cite{BBP98}.
The $hep$ process, as the proton weak capture on $^3$He
is known, is the only source of solar neutrinos
with energies larger than 15 MeV--their end-point energy is
about 19 MeV.  This fact has naturally led
to questions about the reliability of
the currently accepted SSM value for the astrophysical
factor at zero energy, $2.3 \times 10^{-20}$ keV~b~\cite{Sch92}.
In particular, Bahcall and Krastev~\cite{BK98} have shown that
a large enhancement, by a factor in the range 20--30, of
the SSM $S$-factor value given above would essentially fit the observed
excess~\cite{Fuk99} of recoiling electrons.

The theoretical description of the $hep$ process, as well as that
of the neutron and proton radiative captures on deuteron and $^3$He,
constitute a challenging problem from the standpoint of
nuclear few-body theory.  Its difficulty can be appreciated
by comparing the measured values for the cross section
of thermal neutron radiative capture on $^1$H, $^2$H, and $^3$He.
Their respective cross sections are: $334.2 \pm 0.5$ mb~\cite{CWC65},
$0.508 \pm 0.015$ mb~\cite{JBB82},
and $0.055 \pm 0.003$ mb~\cite{Wol89}.  Thus, in going from $A$=2 to 4
the cross section has dropped by almost four orders of magnitude.
These processes are induced by magnetic-dipole transitions between
the initial two-cluster state in relative S-wave and the final bound state.
In fact, the inhibition of the $A$=3 and 4 captures has been understood
for a long time~\cite{Sch37}: the $^3$H and $^4$He states are approximate
eigenstates of the magnetic dipole operator ${\bbox \mu}$, and consequently
matrix elements of $\mu_z$ between $nd$ ($n\,^3$He) and $^3$H ($^4$He) vanish
(approximately) due to orthogonality.  This orthogonality argument fails
in the case of the deuteron, since then $\mu_z$ can connect the large S-wave
component of the deuteron to an isospin $T$=1 $^1$S$_0$ $np$ state.

This quasi-orthogonality, while again
invalid in the case of the proton weak capture on protons~\cite{Sch98},
is also responsible for inhibiting the $hep$ process.  Both these
reactions are induced by the Gamow-Teller operator, which differs
from the (leading) isovector spin part of the
magnetic dipole operator essentially by an isospin
rotation.  As a result, the $hep$ weak capture and $nd$, $pd$ and
$n\,^3$He radiative captures are extremely sensitive to: i) the D-state admixtures
generated by tensor interactions, and ii) many-body terms in the
electroweak current operator.  For example, many-body
current contributions provide, respectively, 50 \% and over 90 \% of the
calculated $pd$~\cite{Viv00} and $n\,^3$He~\cite{Sch92,Car90}
cross sections at very low energies.

In this respect, the $hep$ weak capture is a particularly delicate
reaction, for two additional reasons:
firstly and most importantly, the one- and many-body current contributions
are comparable in magnitude, but of opposite
sign~\cite{Sch92,Car91}; secondly, many-body axial currents, specifically
those arising from excitation of $\Delta$ isobars which
give the dominant contribution, are
model dependent~\cite{Car91}.
This destructive interference between one-
and many-body currents also occurs
in the $n\,^3$He (\lq\lq $hen$\rq\rq)
radiative capture~\cite{Sch92,Car90}, with the difference
that there the leading components of the many-body currents
are model independent, and give a much larger contribution than
that associated with the one-body current.

The cancellation in the $hep$ process between the one-
and two-body matrix elements has the effect of enhancing
the importance of P-wave capture channels.
Indeed, one of the results of the present
work is that these channels give about 40 \% of the
$S$-factor calculated value.  That the $hep$ process could proceed as easily
through P- as S-wave capture was not
not sufficiently appreciated~\cite{WB73} in all
earlier studies of this reaction we are aware of, with the
exception of Ref.~\cite{BK98}, in which Horowitz suggested, on the basis
of a very simple one-body reaction model, that the $^3$P$_0$ channel may
be important.

Most of the earlier studies~\cite{Wol89,WB73,TEG83} had attempted to relate
the matrix element of the axial current occurring in the
$hep$ capture to that of the electromagnetic current
in the $hen$ capture, exploiting (approximate) isospin symmetry.  This
approach led, however, to $S$-factor values ranging from $3.7$
to $57$, in units of $10^{-20}$ keV~b.  In an
attempt to reduce the uncertainties
in the predicted values for both the radiative
and weak capture rates, {\it ab initio}
microscopic calculations of these reactions were performed in the early
nineties~\cite{Sch92,Car90,Car91}, using variational
wave functions corresponding to a realistic
Hamiltonian, and a nuclear electroweak current consisting of one- and
many-body components.  These studies showed
that inferring the $hep$ $S$-factor from the measured $hen$ cross section
can be misleading, because of different
initial-state interactions in the $n\,^3$He and
$p\,^3$He channels, and because of the large contributions
associated with the two-body components of the electroweak
current operator, and their destructive interference with
the one-body current contributions.  

The significant progress made in the
last few years in the modeling of two- and three-nucleon interactions and
the nuclear weak current, and the description of the bound and
continuum four-nucleon wave functions, have prompted
us to re-examine the $hep$ reaction.  In the present work we briefly
summarize the salient points in the calculation, and report our results
for the $S$-factor in the energy range 0--10 keV.  An exhaustive account
of this study~\cite{Mar00}, however, will be published elsewhere.
 
The cross section for the $^{3}$He($p$,$e^{+}\nu_{e}$)$^{4}$He reaction
at a c.m. energy $E$ is written as 

\begin{eqnarray}
\sigma(E)=\int&& 2\pi \, \delta\left (\Delta m  + E -
\frac{q^{2}}{2 m_{4}} - E_e
- E_\nu\right )\frac{1}{v_{\rm rel}} \nonumber \\
&&\times \frac{1}{4}\sum_{s_e s_\nu}\sum_{s_1 s_3}
|\langle f\,|\,H_{W}\,|\,i\rangle|^{2}
\frac{d{\bf{p}}_{e}}{(2\pi)^3} \frac{d{\bf{p}}_{\nu}}{(2\pi)^3} \ ,
\label{eq:xsc1}
\end{eqnarray}
where $\Delta m = m + m_3 - m_4 $ = 19.8 MeV
($m$, $m_3$, and $m_4$ are the proton, $^3$He, and $^4$He rest masses,
respectively), $v_{\rm rel}$ is the $p\,^{3}$He relative velocity, and
the transition amplitude is given by

\begin{equation}
\langle f|H_{W}|i\rangle=
\frac{G_{V}}{\sqrt{2}}\,l^{\sigma} \langle -{\bf{q}}; ^{4}\!{\rm{He}}|
j_{\sigma}^{\dag}({\bf{q}})|{\bf{p}}; p\,^{3}{\rm{He}}\rangle \ .
\label{eq:tra}
\end{equation}
Here $G_V$ is the Fermi constant, ${\bf{q}}={\bf{p}}_{e}+{\bf{p}}_{\nu}$,
$|{\bf{p}}; p\,^{3}{\rm{He}}\rangle$ and
$|-\!{\bf{q}}; ^{4}\!{\rm{He}} \rangle$ represent, respectively,
the $p\,^{3}$He scattering state with relative momentum
${\bf{p}}$ and $^{4}$He bound state recoiling with momentum $-{\bf{q}}$,
$l_{\sigma}$ is the leptonic weak current,
$l_{\sigma} = \overline{u}_{\nu}\gamma_{\sigma}(1-\gamma_5)v_{e}$ (the
lepton spinors are normalized as $v_{e}^{\dag}v_{e}={u}_{\nu}^{\dag}{u}_{\nu}=1$),
and $j^\sigma({\bf q})$ is the nuclear weak current, $j^\sigma({\bf q})
=( \rho({\bf q}), {\bf j}({\bf q}))$.  The dependence of the amplitude
upon the spin projections of the leptons, proton and $^3$He has been
omitted for ease
of presentation.  Since the energies of interest are of the order
of 10 keV or less--the Gamow-peak energy is 10.7 keV for the $hep$
reaction--it is convenient to expand the $p\,^3$He scattering state
into partial waves, and perform a multipole decomposition of the nuclear
weak charge, $\rho({\bf q})$, and current, ${\bf j}({\bf q})$, operators.
Standard manipulations lead to~\cite{Mar00}

\begin{equation}
\frac{1}{4}\sum_{s_e s_\nu}\sum_{s_1 s_3}
|\langle f\,|\,H_{W}\,|\,i\rangle|^{2} = (2 \pi)^2\> G_V^2\> L_{\sigma \tau} \>
N^{\sigma\tau} ,
\end{equation}
where the lepton tensor $L^{\sigma \tau}$ is written in terms of
electron and neutrino four-velocities
as $L^{\sigma \tau}={\rm v}_{e}^{\sigma}{\rm v}_{\nu}^{\tau}+
{\rm v}_{\nu}^{\sigma}{\rm v}_{e}^{\tau}-g^{\sigma\tau}
{\rm v}_{e} \cdot {\rm v}_{\nu}+
{\rm{i}}\>\epsilon^{\sigma\alpha\tau\beta}{\rm v}_{e,\alpha} {\rm v}_{\nu,\beta}$, while
the nuclear tensor is defined as

\begin{equation}
N^{\sigma \tau} \equiv \sum_{s_1 s_3} W^\sigma({\bf q};s_1 s_3)
W^{\tau *}({\bf q}; s_1 s_3) \ ,
\label{nuclt}
\end{equation}
with
\begin{eqnarray}
W^{\sigma=0,3}({\bf q}; s_1 s_3)&=&\sum_{LSJ} X^{LSJ}_0(\hat{\bf q};s_1 s_3)
 T^{LSJ}_J(q) \ , \\
W^{\sigma=\lambda}({\bf q}; s_1 s_3)&=&
-\frac{1}{\sqrt{2}} \sum_{LSJ} X^{LSJ}_{-\lambda}(\hat{\bf q};s_1 s_3) \nonumber \\
& &\left [\lambda \, M^{LSJ}_J(q) + E^{LSJ}_J(q) \right ] \ ,
\end{eqnarray}
where $\lambda = \pm 1$ denote spherical components.
The functions $X_{\lambda=0,\pm 1}$ depend upon the direction
$\hat{\bf q}$, and the proton and $^3$He 
spin projections $s_1$ and $s_3$~\cite{Mar00}
(note that the quantization axis for the
hadronic states is taken along $\hat{\bf p}$, the direction
of the $p\,^3$He relative momentum), while
$T^{LSJ}_J$=$C_{J}^{LSJ}$ or $L_{J}^{LSJ}$ for $\sigma$=0 or 3.
The quantities $C_J^{LSJ}$, $L_J^{LSJ}$,
$E_{J}^{LSJ}$ and $M_{J}^{LSJ}$ are the reduced matrix
elements (RMEs) of the Coulomb $(C_{\ell \ell_z})$,
longitudinal $(L_{\ell \ell_z})$,
transverse electric $(E_{\ell \ell_z})$,
and transverse magnetic $(M_{\ell \ell_z})$
multipole operators between the initial $p\,^3$He state
with orbital angular momentum $L$, channel spin $S$ ($S$=0,1),
and total angular momentum $J$, and final $^4$He state.
The present study includes S- and P-wave capture channels,
i.e. the $^1$S$_0$, $^3$S$_1$, $^3$P$_0$, $^1$P$_1$, $^3$P$_1$, and $^3$P$_2$
states, and retains all contributing
multipoles connecting these states to the $J^\pi$=0$^+$ ground
state of $^4$He.

The bound- and scattering-state wave functions are obtained variationally
with the correlated-hyperspherical-harmonics (CHH) method, developed in
Refs.~\cite{VKR95,VRK98}.  The nuclear Hamiltonian
consists of the Argonne $v_{18}$ two-nucleon~\cite{WSS95} and Urbana-IX
three-nucleon~\cite{Pud95} interactions.  This realistic
Hamiltonian, denoted as AV18/UIX,
reproduces the experimental binding energies and charge
radii of the trinucleons and $^4$He in \lq\lq exact\rq\rq Green's function
Monte Carlo (GFMC) calculations~\cite{Pud97}.  The binding energy of $^4$He
calculated with the CHH method~\cite{Mar00,VKR95}
is within 1 \% of that obtained with the GFMC
method.  The accuracy of the CHH method to
calculate scattering states has been successfully
verified in the case of three-nucleon systems, by comparing results for
a variety of $Nd$ scattering observables
obtained by a number of groups using different
techniques~\cite{Kie98}.  Studies along similar lines~\cite{Viv98}
to assess the accuracy of the CHH solutions for the
four-nucleon continuum have already begun.

The CHH predictions~\cite{VRK98} for the
$n\,^3$H total elastic cross section and coherent scattering
length have been found to be in excellent agreement with the corresponding
experimental values.  The $n\,^3$H cross section is known over
a rather wide energy range, and its extrapolation to zero energy
is not problematic~\cite{Phi80}.  The situation is different for the $p\,^3$He
channel, for which the singlet and triplet scattering lengths
$a_{\rm s}$ and $a_{\rm t}$ have been determined
from effective range extrapolations of data taken above 1 MeV, and
are therefore somewhat uncertain, $a_{\rm s}=(10.8 \pm 2.6)$ fm~\cite{AK93} and
$a_{\rm t}=(8.1 \pm 0.5)$ fm~\cite{AK93}
or $(10.2 \pm 1.5)$ fm~\cite{TEG83}.  Nevertheless,
the CHH results are close to the experimental values above:
the AV18/UIX Hamiltonian predicts~\cite{VRK98}
$a_{\rm s}=11.5$ fm and $a_{\rm t}=9.13$ fm.  At low
energies (below 4 MeV) $p\,^3$He elastic scattering
proceeds mostly through S- and P-wave channels, and the CHH
predictions, based on the AV18/UIX model, for the differential
cross section~\cite{VKR00} are in good agreement
with the experimental data.

The nuclear weak current has vector and axial-vector parts, with
corresponding one- and many-body components.  The one-body
components have the standard expressions obtained from a non-relativistic
reduction of the covariant single-nucleon vector and axial-vector
currents, including terms proportional to $1/m^2$.
The two-body weak vector currents
are constructed from the isovector two-body 
electromagnetic currents in accordance with the conserved-vector-current
hypothesis, and consist~\cite{Mar00} of \lq\lq model-independent\rq\rq
and \lq\lq model-dependent\rq\rq terms.  The model-independent
terms are obtained from the nucleon-nucleon interaction, and by
construction satisfy current conservation with it.
The leading two-body weak vector current
is the \lq\lq$\pi$-like\rq\rq operator, obtained from the isospin-dependent
spin-spin and tensor nucleon-nucleon interactions.
The latter also generate an isovector \lq\lq$\rho$-like\rq\rq current,
while additional isovector two-body currents arise from the isospin-independent
and isospin-dependent central and momentum-dependent interactions.  These
currents are short-ranged, and numerically far less important
than the $\pi$-like current.  With the exception of the $\rho$-like current,
they have been neglected in the present work.  The model-dependent
currents are purely transverse, and therefore cannot be directly
linked to the underlying two-nucleon interaction.  The present
calculation includes the currents associated with excitation
of $\Delta$ isobars which, however, are found to give a rather small
contribution in weak-vector transitions, as compared to that due to
the $\pi$-like current.  The $\pi$-like and $\rho$-like contributions
to the weak vector charge operator~\cite{Mar00} have also been retained in the
present study.

The leading many-body terms in the
axial current due to $\Delta$-isobar
excitation are treated non-perturbatively in the 
transition-correlation-operator (TCO) scheme, originally
developed in Ref.~\cite{Sch92}
and further extended in Ref.~\cite{Mar98}.  In the
TCO scheme--essentially, a scaled-down approach to
a full $N$$+$$\Delta$ coupled-channel treatment--the
$\Delta$ degrees of freedom are explicitly
included in the nuclear wave functions.  The axial charge operator includes, in
addition to $\Delta$-excitation terms (which, however,
are found to be unimportant~\cite{Mar00}),
the long-range pion-exchange term~\cite{Kub78}, required by low-energy
theorems and the partially-conserved-axial-current relation,
as well as the (expected) leading short-range terms constructed from the central
and spin-orbit components of the nucleon-nucleon interaction, following
a prescription due to Riska and collaborators~\cite{Kir92}.

The largest model dependence is in the weak axial current.  To minimize it,
the poorly known $N$$\Delta$ transition
axial coupling constant $g_A^*$ has been
adjusted to reproduce the experimental value
of the Gamow-Teller matrix element in tritium
$\beta$-decay~\cite{Sch98,Mar00}.  While this
procedure is model dependent, its actual model dependence is in fact
very weak, as has been shown in Ref.~\cite{Sch98}.

The calculation proceeds in two steps~\cite{Mar00}: first, the matrix elements
of $\rho({\bf q})$ and ${\bf j}({\bf q})$ between
the initial $p\,^3$He $LSJJ_z$ states and final $^4$He are
calculated with Monte Carlo integration techniques; second,
the contributing RMEs are extracted from these matrix elements, and
the cross section is calculated by performing the integrations
over the electron and neutrino momenta in Eq.~(\ref{eq:xsc1})
numerically, using Gauss points.

The results for the $S$-factor, defined as $S(E) = E\, \sigma(E)\,
{\rm exp}( 4\, \pi \, \alpha/v_{\rm rel})$ ($\alpha$ is the fine
structure constant), 
at $p\,^3$He c.m.\ energies of $0$, $5$, and $10$ keV
are reported in Table~\ref{tb:sfact}.  In the table, the column
labelled S includes both the $^1$S$_0$ and $^3$S$_1$ channel 
contributions, although the former are at the level of a few parts
in $10^{3}$.  The energy dependence is rather weak:
the value at $10$ keV is only about 4 \% larger
than that at $0$ keV.  The P-wave capture states are found to
be important, contributing about 40 \% of the calculated
$S$-factor.  However, the contributions from D-wave channels 
are expected to be very small.  We have verified explicitly
that they are indeed small in $^3$D$_1$ capture.
The many-body axial currents associated with $\Delta$ excitation play a crucial
role in the (dominant) $^3$S$_1$ capture, where they reduce
the $S$-factor by more than a factor of four.  Thus
the destructive interference between the one- and many-body
current contributions, first obtained in Ref.~\cite{Car91}, is
confirmed in the present study.  The (suppressed) one-body contribution
comes mostly from transitions involving the D-state
components of the $^3$He and $^4$He wave functions, while
the many-body contributions are predominantly due to transitions
connecting the S-state in $^3$He to the D-state in $^4$He, or viceversa.

It is important to stress the differences between the present
and all previous studies.  Apart from
ignoring, or at least underestimating, the contribution
due to P-waves, the latter only considered the long-wavelength
form of the weak multipole operators, namely, their $q$=$0$ limit.
In $^3$P$_0$ capture, for example, only the $C_0$-multipole,
associated with the weak axial charge, survives in this limit,
and the corresponding $S$-factor is calculated to be $2.2 \times 10^{-20}$
keV~b, including two-body contributions.  However,
when the transition induced by the longitudinal component of
the axial current (via the $L_0$-multipole, which
vanishes at $q$=$0$) is also taken into account, the $S$-factor
becomes $0.82 \times 10^{-20}$ keV~b, because of a destructive
interference between the $C_0$ and $L_0$ RMEs.  Thus use
of the long-wavelength approximation in the calculation of the
$hep$ $S$-factor leads to inaccurate results.
 
Finally, besides the differences listed above, the present
calculation also improves that of Ref.~\cite{Sch92} in a number of
other important respects: firstly, it uses accurate
CHH wave functions, corresponding to the last generation
of realistic interactions; secondly,
the model for the nuclear weak current has been extended to include
the axial charge as well as the vector charge and current operators.
Thirdly, the one-body operators now take
into account the $1/m^2$ relativistic
corrections, which were previously neglected.  In $^3$S$_1$
capture, for example, these terms increase by 25 \% the
dominant (but suppressed) $L_1$ and $E_1$ RMEs calculated with the
(lowest order) Gamow-Teller operator.  These improvements in the
treatment of the one-body axial current indirectly affect also the contributions
of the $\Delta$-excitation currents~\cite{Mar00}, because of the
procedure used to determine the coupling constant $g_A^*$. 

To conclude, we have carried out a realistic calculation
of the $hep$ reaction, predicting a value for the 
$S$-factor five times larger than that
used in the SSM.  This enhancement, while very significant,
is far smaller than that required by fits to the Super-Kamiokande data.
Although the present result is inherently model dependent, it
is unlikely the model dependence be so large to accomodate
a drastic increase in the prediction obtained here.
 
The authors wish to thank J.F.\ Beacom, J.\ Carlson, V.R.\ Pandharipande,
D.O.\ Riska, P.\ Vogel, and R.B.\ Wiringa for useful discussions.
The support of the U.S. Department of Energy under contract number
DE-AC05-84ER40150 is gratefully acknowledged by L.E.M.\ and R.S. 
\begin{table}
\caption{The $hep$ $S$-factor, in units of $10^{-20}$ keV~b, calculated
with CHH wave functions corresponding to the AV18/UIX Hamiltonian model,
at $p\,^3$He c.m.\ energies $E$=$0$, $5$, and $10$ keV.  The rows labelled
\lq\lq one-body\rq\rq and \lq\lq full\rq\rq list the contributions
obtained by retaining the one-body only and both one- and many-body
terms in the nuclear weak current.  The contributions due to the
S-wave channels only and S- and P-wave channels are
listed separately.  The Monte Carlo statistical error is 
at the 5\% level on the total $S$-factor.}
\begin{tabular}{ccccccc}
$E$ (keV) & \multicolumn{2}{c} {$0$} & 
            \multicolumn{2}{c} {$5$} & 
            \multicolumn{2}{c} {$10$} \\
\tableline
& S & S+P & S & S+P & S & S+P\\
\tableline
one-body  &26.4  & 29.0 & 25.9 & 28.7 & 26.2 & 29.3 \\
full      &6.39  & 9.64 & 6.21 & 9.70 & 6.37 & 10.1
\end{tabular}
\label{tb:sfact}
\end{table}
\end{document}